%% file: main.tex
\documentclass[runningheads]{llncs}
\usepackage{graphicx}
\usepackage{csvsimple}
\usepackage{datatool}
\usepackage{multirow}
\newcommand\Tstrut{\rule{0pt}{2.6ex}}         %
\usepackage{soul}
\usepackage{times}  
\usepackage{url}

\usepackage{hyperref}
\usepackage{balance}
\usepackage{xcolor}
\usepackage{comment}
\usepackage[misc,geometry]{ifsym} 
\newcommand{\squishlist}{
 \begin{list}{$\bullet$}
  { \setlength{\itemsep}{0pt}
     \setlength{\parsep}{1pt}
     \setlength{\topsep}{1pt}
     \setlength{\partopsep}{0pt}
     \setlength{\leftmargin}{1.5em}
     \setlength{\labelwidth}{1em}
     \setlength{\labelsep}{0.5em} } }
 \newcommand{\squishend}{\end{list}}

 \usepackage{booktabs}

\begin{document}
\title{You Get What You Chat: Using Conversations to Personalize Search-based Recommendations}
\titlerunning{Using Conversations to Personalize Search-based Recommendations}
\author{Ghazaleh H. Torbati\textsuperscript{(\Letter)},
Andrew Yates, 
Gerhard Weikum}
\authorrunning{G. Torbati, A. Yates and G. Weikum}
\institute{Max-Planck Institute for Informatics\\
Saarland Informatics Campus,
Saarbrücken, Germany\\
\email{\{ghazaleh,ayates,weikum\}@mpi-inf.mpg.de}}
\maketitle
\pagestyle{plain}
\begin{abstract}
Prior work on personalized recommendations has focused on exploiting explicit signals
from user-specific queries, clicks, likes, and ratings.
This paper investigates tapping into a different source of implicit signals of interests and tastes: online chats between users. The paper develops an
expressive model and effective methods for
personalizing search-based entity recommendations.
User models derived from chats 
augment different methods for
re-ranking entity answers for medium-grained queries.
The paper presents specific techniques to enhance the user models by capturing
domain-specific vocabularies and by
entity-based expansion.
Experiments are based on a 
collection of
online chats from a controlled user study covering three domains: books, travel, food.
We evaluate different configurations
and compare
chat-based user models 
against concise
user profiles from questionnaires.
Overall, these two variants perform on par in terms
of NCDG@20, but each has advantages in certain domains. 

\end{abstract}
\input{chapters/introduction}
\input{chapters/methodology}

\input{chapters/experiments}

\input{chapters/relatedwork}

\input{chapters/conclusion}

\subsubsection{Acknowledgements.}
This research was supported by the ERC Synergy Grant 610150 (imPACT).
\bibliographystyle{splncs04}
\bibliography{ref}
\end{document}

%% file: chapters/introduction.tex
\section{Introduction}

\noindent{\bf Motivation:}
Recommender systems are at the heart of {\em personalized} shopping and
online services for music and video streaming, hotels and restaurants, or food recipes \cite{DBLP:reference/sp/2015rsh,DBLP:conf/sigir/BalogRA19,DBLP:journals/corr/abs-2003-00911}.
{\em Search-based recommendation} is a setting where the user starts with
a query and the recommendation model determines the result ranking based
on the user's interests and preferences.
This paper considers
medium-grained queries about product entities
(books, food recipes, and travel destinations) such as 
{\em paranormal romance} or {\em wine lover destinations} -- in contrast to coarse-grained queries such as
{\em love novels} or {\em Europe} and fine-grained queries such as 
{\em similar to Stephenie Meyer's Twilight} or {\em vineyards of the Bourgogne}.
 Results are assumed to come from a %
 search engine
(restricted to suitable domains for the respective vertical). 
Therefore, the personalization amounts to {\em re-ranking}
the top results with regard to a model of the user's individual tastes.

For this setting, the {\em user model} or {\em profile} can be represented
explicitly in a personal knowledge base \cite{DBLP:conf/ictir/BalogK19} or implicitly in a
latent model \cite{koren2009matrix,DBLP:journals/csur/ZhangYST19}. 
These models can be constructed from various kinds of observations
on user behavior:
\squishlist
\item[A:] Explicit signals like clicks, likes, ratings and purchases.
\item[B:] %
User profiles such as
{\small\url{adssettings.google.com}} where users can see and check or un-check topics (even if the profile itself is learned from other signals).
\item[C:] Implicit signals from other online behavior, like social media
posts or conversations with other users.
\squishend
Option A is most widely used in practice 
(e.g., \cite{DBLP:conf/recsys/ZhaoHWCNAKSYC19,Lalmas:NetflixWorkshop,DBLP:conf/www/JiangWRCYCGJC020})
and includes standard recommenders
based on collaborative filtering \cite{sarwar2001item}.
However, this
rich kind of data is available only to major service providers,
such as music streaming where playlists and other I-like-the-song
signals are abundant.
Option B operates on concise digests of user interests and item properties, for example,
a list 
of topics and 
tags
(e.g.,  \cite{DBLP:conf/sigir/WuG20}).
This is less 
informative than A, but has the advantage that the user can easily
interpret her profile and adjust it at her discretion (e.g., dropping a topic that
is unwanted). 
Option C has been studied for recommending news and discussions, 
but the best signals are still the
user histories of
clicks, dwell times and likes (e.g., \cite{DBLP:conf/iui/LiuDP10,DBLP:conf/acl/WuQCWQLLXGWZ20}).
For search-based recommendation of product entities, C has not been explored at all,
except for the specific case of leveraging product reviews (e.g., \cite{DBLP:journals/umuai/ChenCW15,DBLP:conf/kdd/Bauman0T17,SIGIR2020:sachdeva:useful}).

This paper focuses on option C. 
It investigates how
online chats between users can be leveraged for personalization in the outlined setting.
To the best of our knowledge, it is the first work that studies chats as a source
for search-based recommendation.

\vspace*{0.1cm}
\noindent{\bf Research Questions:}
We investigate the following research questions:
\squishlist
\item RQ0: How can we leverage signals from {\em user-user chats} to personalize search-based recommendations across a {\em variety of domains}: books, food recipes, and travel destinations?
\item RQ1: How do methods that tap into individual {\em conversations} compare to methods that merely access {\em concise user profiles}?
\item RQ2: How important is it to {\em customize} the per-user models to the {\em specific domain} at hand, for example, books vs. travel?
\item RQ3: How much added value can we get from {\em entity awareness}: detecting named entities in user chats, mapping them to a background knowledge base, and using that information for expansion of user models and re-ranking techniques?
\squishend

\vspace*{0.1cm}
\noindent{\bf Contributions:}
We devise techniques for constructing language models and using them for re-ranking,
with various components derived from chats: i) computing domain-specific vocabularies and
ii) entity detection and entity-based expansions. 
The chats are recorded real-time conversations, gathered in a substantial
user study
with 
14
students and 
83
pair-wise chats 
(with 9,797 utterances and 59k
tokens in total and a total duration of 93 hours).
We contrast chat-based personalization against techniques that merely build on concise user profiles derived from short questionnaires \cite{CHIIR20:PES}.
The paper makes the following contributions:
\squishlist
\item It is the first approach to consider user chats as a source for search-based recommendation across a variety of vertical domains. Chats are a rich source of information about individual interests and tastes. In contrast to latent models learned from clicks, likes, ratings, etc., a user can more easily interpret and edit/censor this information to selectively restrict its usage for privacy reasons.
\item We systematically compare chat-based personalization against a more restrictive approach that merely uses concise user profiles based on short questionnaires.
In our experiments, both show
advantages in certain domains, and perform
on par overall.
\item We devise techniques for per-domain customization by controlling the vocabulary and appropriate weighting of terms, and report on their experimental effectiveness.
\item  We devise techniques to harness entities and background knowledge in the construction of user models, and report on their experimental effectiveness.
\item We release a dataset consisting of filled questionnaires, 
pair-wise user chats, document URLs, and search result assessments by users for three domains (books, travel, food).
The data is available at
{\small\url{http://personalization.mpi-inf.mpg.de/}}
\squishend

%% file: chapters/methodology.tex
\section{Computational Model and Re-ranking}
\label{sec:methodsandmodels}
We approach 
the personalization of entity-search 
answers by
re-ranking a pool of initial non-personalized results
using three different methods for scoring and ranking: the BM25 family,
statistical language models, and neural ranking.
Beginning with these ranking methods, we incorporate a user model to personalize results and
domain-specific term weights to identify terms that are informative with a domain.
We additionally apply expansion techniques to expand entities found in the user model.
Rerankers thus consider a user model in addition to queries and documents. In our setting,

\squishlist
\item {\bf Queries} are short, medium-grained bags of words (or phrases),
such as ``scandinavian suspense'' (for the books domain) or
``wine lover destinations'' (travel).%
\item {\bf Documents} are entity-level answers obtained from specific websites that provide comprehensive contents about three domains: 
{\small\url{goodreads.com}} for books,
{\small\url{wikivo-yage.org}} for travel,
and {\small\url{allrecipes.com}} for food.
Each answer has a key entity that can be
easily identified (e.g., from the URL string
or page title) and comprises an informative
description of the entity.
Two of the sites include also 
extensive reviews and discussion by their
communities.
\item {\bf User models} represent a user's interests and tastes as a bag of words (or n-grams) taken from either a short
{\em questionnaire/profile} filled in by the user or
a {\em collection of online chats} with other users.
Both of these options are further refined by instructing users to focus on specific scopes:
general,
books, 
travel, and food.
This yields 8 basic options for the user model, which we further augment with techniques for domain-specific vocabularies and entities.
\squishend

\begin{figure}[h]
\centering
    \includegraphics[width=\linewidth]{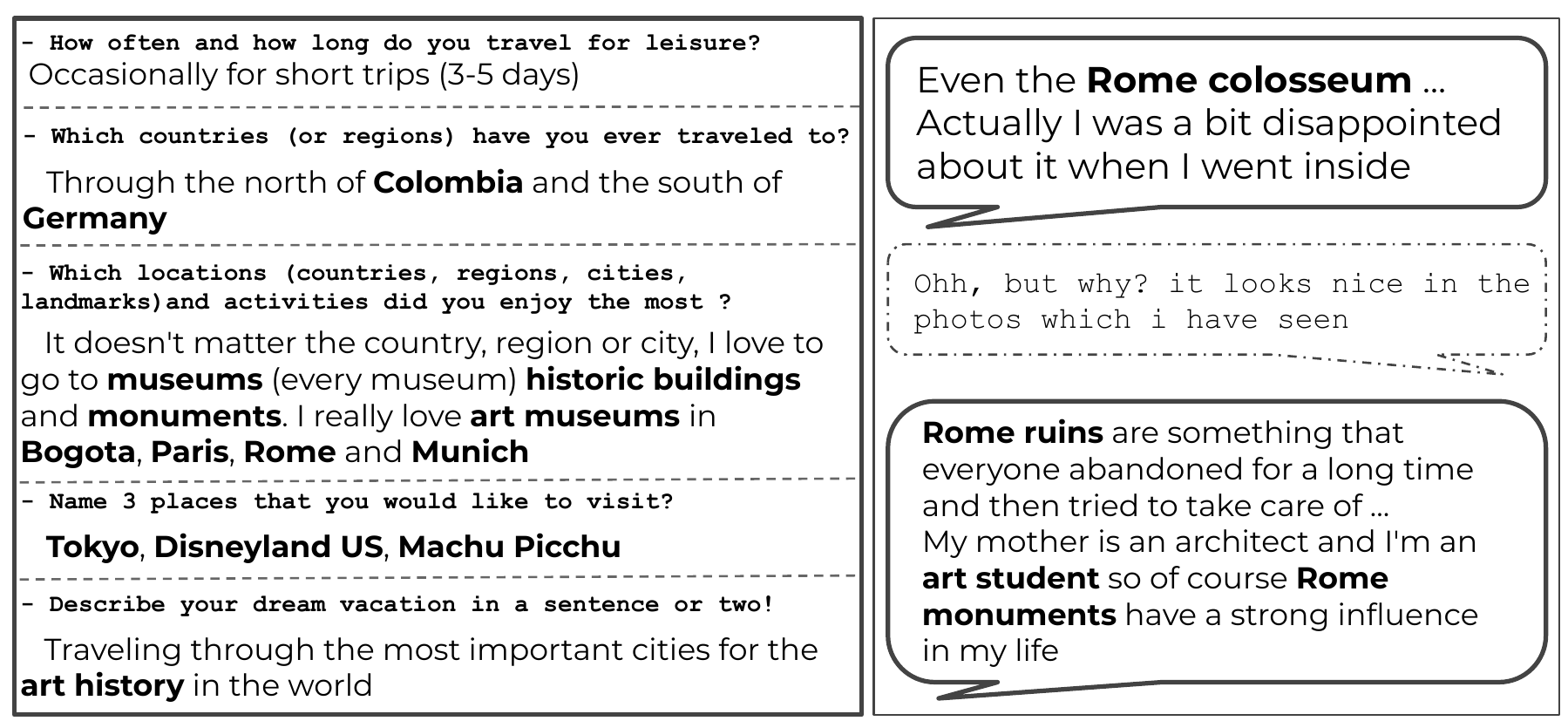}
\vspace*{-0.5cm}
\caption{Excerpts from user questionnaire and chat on travel domain (with recognized named entities and concepts in boldface)}
    \label{fig:usermodelsexample}
\end{figure}

\noindent For illustration,
Figure \ref{fig:usermodelsexample} shows excerpts from
the questionnaire and the chat collection for an example user. For the query ``temples and culture'', this
user-specific information led to high ranks of
travel destinations like Borobudur, Delphi and Ellora
-- all confirmed as very good recommendations by that user.

\subsection{Re-Ranking Methods}

Given a query {\em q},  a user model {\em u}, and a document {\em d} from a pool of non-personalized results,
we personalize the results by re-ranking them according to the user model.
We explore three re-ranking methods for doing this.

\noindent{\bf Language Models:} 
The first variant for re-ranking is based on
language models (LMs) \cite{DBLP:journals/ftir/Zhai08},
which provide a natural way to incorporate the user model.
We compute the Kullback-Leibler divergence between a query model and a document model
with Dirichlet
smoothing over unigrams or n-grams.
In pilot experiments, unigrams outperformed bigrams and trigrams;
hence we focus on the unigram case.
To personalize for a specific user, we compute
the Kullback-Leibler divergence i) between 
the query $q$ and the document $d$
and ii) between 
the user model $u$ and the document $d$.
These two components are combined
into a linear mixture with
hyper-parameter $\lambda$.
Additionally, we incorporate a background model
$C$ for smoothing, based on ClueWeb'09. That is,

\begin{equation}
\begin{split}
score(q,d,u) \propto - (\lambda div(\theta_q\| \theta_{d}) + (1-\lambda) div(\theta_u\| \theta_{d})) \propto \\
- \lambda \sum_{w \in V_q} \text{spy}(w) \cdot
  p(w|\theta_q) \log \frac{p(w|\theta_q)}{(p(w|\theta_d)+ \mu p(w|\theta_C)) ~/~ (|d|+\mu)}\\
- (1-\lambda) \sum_{w \in V_u} \text{spy}(w) \cdot
  p(w|\theta_u) \log \frac{p(w|\theta_u)}{(p(w|\theta_d)+ \mu p(w|\theta_C)) ~/~ (|d|+\mu)} 
\end{split}
\label{eq2}
\end{equation}
where 
$V_u$ and $V_q$ are the vocabularies of the 
user and query models, and
$\theta_q$, $\theta_u$,
$\theta_d$ and $\theta_C$ denote the multinomial parameters of query, user, document and background models, with Dirichlet smoothing parameter
$\mu$ set to the average document length.
We 
introduce 
additional weights
$\text{spy(t)}$
which reflect the
specificity of a term $t$ for a given domain
(books, food or travel),
as described in Section \ref{sec:domainspecificity}.
This can be viewed as conditioning the query and user models with a domain model.

Optionally, we integrate word embeddings by using
the cosine distance between precomputed word2vec
embeddings \cite{DBLP:conf/nips/MikolovSCCD13} as a term-term similarity score
$sim(w,t)$.
This is plugged into the document model
by means of a translation model largely following \cite{DBLP:conf/cikm/KuziSK16},
with per-term contributions 
$p(w|\theta_d)$ replaced by 
summing over all similar terms (above a threshold): 
$\sum\limits_{t: w \sim t}^{} sim(w,t) \cdot p(t|\theta_d)$. %

\vspace*{0.2cm}
\noindent{\bf BM25:} 
The second variant for re-ranking is the Okapi BM25 model \cite{DBLP:journals/ftir/RobertsonZ09}.
We incorporate the user model by query expansion.
In principle,
all terms from the entire
chat collection of a user are added to the query. 
We will discuss 
ways of reducing noise
and focusing the query in Sections
\ref{sec:domainspecificity} and \ref{sec:entityexpansion}. That is,

\begin{equation}
score(q \cup u,~d) \propto \sum _{w \in V_{q \cup u}} {\text{spy}(w)} \cdot \text{idf}(w)\cdot {\frac {\text{tf}(w, d) \cdot (k_{1}+1)}{\text{tf}(w, d) +k_{1}\cdot \left(1-b+b\cdot {\frac {|d|}{\text{avgdl}}}\right)}} 
\label{eq1}
\end{equation}
with domain-specificity weight $\text{spy(w)}$, document length $|d|$, average
document length $avgdl$, and BM25
parameters $b$ and $k_1$.

\vspace*{0.2cm}
\noindent{\bf Neural Ranking with KNRM:}
The third variant for re-ranking 
is the KNRM neural 
method~\cite{DBLP:conf/sigir/XiongDCLP17}
which takes a bag-of-words query as input.
KNRM produces a query-document relevance score by comparing embedding similarities between query and document terms.
During training, KNRM learns how to weigh different embedding similarity levels.
As with BM25, we incorporate the user model by query expansion.

\section{Domain-specific Vocabulary Weighting}
\label{sec:domainspecificity}
As described in the previous section, the ranking models are further augmented 
by awareness of domain-specific
vocabularies, customizing the user models
and document models
to books, travel or food, respectively.
The intuition is that terms in a user chat
are informative if they refer to a certain meaning
within a particular domain.
For example, terms like ``history'' or
``museum'' are good cues about a user's travel interests,
whereas terms like ``price'' or ``bargain'' are
uninformative -- although all these terms 
have
comparable idf values in large corpora.

We incorporate this idea of domain specificity
by computing per-domain weights for terms,
and weighing term contributions by the various %
ranking models accordingly (or even eliminating
low-weight terms).
To this end, we estimate the conditional probability
of a term occurring in a domain-specific context
(document or chat) given that it occurs in a general corpus:
\begin{equation}
\text{spy}(w) = P(w \in Dom | w \in All) %
\propto
\frac{tf(w \in Dom) / |Dom|}{tf(w \in All) / |All|}
\end{equation}

As underlying text collections for this estimator,
we use the pool of all retrieved documents per domain
(e.g., all answers for book search, including
book descriptions and user reviews) against 
the pool of documents for all three domains together. 
We also experimented with 
term weighting for user-specific vocabularies,
contrasting all chats by the same user
against 
a universal corpus.
This 
did not lead to
significant changes in the empirical results, though,
and is disregarded in the following.

\section{Entity Expansion}
\label{sec:entityexpansion}

\noindent{\bf Named Entity Recognition and Disambiguation (NER/NED):}
Among all terms and phrases 
in the user's chats and questionnaires,
entities and concepts deserve specific treatment.
We ran standard NER ({\small\href{https://stanfordnlp.github.io/CoreNLP/}{stanfordnlp.github.io}})
and NED ({\small\href{https://github.com/ambiverse-nlu}{github.com/ambiverse-nlu}}) tools
to link text spans to uniquely
identified entities in the YAGO
knowledge base, which in turns links
most of these to Wikipedia.
However, the NER stage 
produced both many false positives and false negatives.
This is largely caused by the very colloquial nature of user chats, with short-hands, misspellings, ungrammatical utterances 
and ad-hoc choice between upper-case and lower-case.
To mitigate this effect, we hired
crowdsourcing workers to mark up
text spans for entities and also 
general concepts that exist in YAGO and Wikipedia (e.g., ``history``
or ``Buddhist art'').
This way we eliminated nearly all
NER errors. As a result, the NED
stage performed well, with precision
reaching approximately 0.83 (estimated
by sampling).
We obtained this perfect mark-up
only for NER as this is much easier 
for crowd workers than NED.

\vspace*{0.2cm}
\noindent{\bf User Model Expansion:}
Rather than adding the names of these detected and linked entities to the user model directly,
which is likely to overfit given that we deal with many long-tail entities (e.g., lesser-known books or special travel destinations), we experimented with expanding entities using embeddings and Wikipedia descriptions.
We first conducted pilot experiments with entity
embeddings using Wikipedia2vec
\cite{DBLP:conf/conll/YamadaS0T16,DBLP:journals/corr/abs-1812-06280} to achieve proper
generalization, but this did not
perform well: many terms that
are highly related by Wikipedia2vec
are quite uninformative if
not misleading 
(e.g., history being most related to
literature; modern, natural,
and wine being most related to coffee, beer, food).
Ultimately, to avoid this noise and topical drift,
we expanded the entities using
their descriptions from (the first
paragraph of) their Wikipedia
articles.
This captures, for example,
content sketches of books,
highlights of travel destinations, etc.
The resulting terms were added
to general as well as domain-specific 
user models.
For the latter, we computed the
domain specificity of an entity
and its descriptive terms, using
the weighting model of
Section \ref{sec:domainspecificity}.

\vspace*{0.2cm}
\noindent{\bf Selective Expansion by Domain-Specificity:}
Some of the extracted entities 
may be poor cues for a certain target domain (e.g. a user chatting about ``Italian cuisine'' is not helpful for books
and could even be misleading for travel).
To counter this potential dilution,
we use the domain-specificity of entities
to filter the candidate entities
before expanding the user model.

To this end, we construct a domain
model for each of the three domains
using Wikipedia2vec embeddings which
capture both entity-level linkage and
textual descriptions
\cite{DBLP:conf/conll/YamadaS0T16,DBLP:journals/corr/abs-1812-06280}.
Candidate entities are mapped into
the same latent space, and the
cosine similarity between entity and domain
is used to select entities above a threshold.
Specifically, the domain vectors
are computed by a weighted average
of the $m=50$  words 
and entities
that are most related to the 
Wikipedia articles on
``book'', ``travel'' and ``food'',
respectively, with weights proportional
to cosine between vectors. 
For selective entity expansion of 
per-domain user models, we pick
entities whose similarity to the respective domain model is above a specified threshold.

This approach introduces several thresholds and hyper-parameters: per-domain numbers of related terms for the domain model and similarity thresholds for pruning entities.
We tuned these via grid search with the objective of maximizing the area under the precision-recall curves for entity detection and disambiguation. We used the 
manually annotated entities in the domain-specific questionnaires
as ground-truth for domain relatedness.

\section{Data Collection}
\label{sec:datacollection}
We gathered personal data in a 4-week user study with 14 
students who were paid
ca. 10 Euros per hour.
We randomly paired two users for 3 chats per week.
For the first week, users were instructed to
chat generally, like mutual introductions.
During the remaining weeks users were asked to chat about specific topical domains: 
users' interests and tastes in books and their experience and interests in
traveling and food. 
On average, each user had 2.8 sessions for each domain, totaling to ca. 11 sessions overall, 
with an average of 653 utterances and 3934 tokens per user.
In addition, each user filled in several questionnaires upfront: a general one with
18 questions about demographics, general interests and personality,
and one for each
of the themes books, travel and food with 2, 5 and 10
questions, respectively (see left side of
Figure \ref{fig:usermodelsexample} for
an example excerpt).
The general questionnaire included personality-oriented questions such as 
``What are your hobbies?'', ``What makes you happy?'', and ``Your golden rule?''.

%% file: chapters/experiments.tex
\section{Experimental Studies}

\subsection{Setup}

The 14 users from whom we collected questionnaire and chat data also 
participated in an assessment study
of personalized search results.
To this end, we compiled 75 medium-grained keyword queries (25 per domain).
Example queries are shown in Table \ref{tab:examplequeries}.

\begin{table}[b]
    \caption{Example queries by domain}
    \vspace*{-0.3cm}
    \centering
    \begin{tabular}{|l|l|l|l|}\hline
Books~ & Scandinavian suspense~~ & Novels made into movies~~ & Personal development~~ \\ \hline
Travel~ & Weekend trip for festival~~ & Best wine lover destination~~ & Epic road trip~~ \\ \hline
Food~ & Perfect breakfast~~ & Iron rich vegetarian recipes~~ & 15-minute meal recipes~~ \\ \hline
    \end{tabular}
    \label{tab:examplequeries}
\end{table}

All queries were issued to a commercial search engine with site restrictions as described in document models (section \ref{sec:methodsandmodels}).
The top-100 answers were retrieved, 
keeping only those that were about
specific entities and discarding 
general list pages -- this left us with 90 or more answers for each query.

The users were asked to identify around 5 queries for each domain on topics that looked
potentially appealing %
to them. This way we avoided personalized judgements on topics that the user does not care about.
For each query, a user assessed 
20 results that were sampled uniformly
at random (to avoid ranking bias)
and, additionally, the top-10 results
from the original ranking (with the risk of bias). 
We asked for subjective, graded assessments
with labels:
2 = strongly interested, 1 = mildly interested, 0 = uninterested, and
discarded all ``I don't know'' assessments.
We required the users to enter justification sentences along with their judgements.
In total, we obtained 
2673
individual assessments for
113 user-query pairs
with 73 distinct queries.

\vspace*{0.2cm}
\noindent{\bf Evaluation Metrics:}
The primary metric is {\bf NDCG@20},
which we use to refer to methods' effectiveness when re-ranking
the 20 randomly sampled query results.
In addition, we report on 
{\bf precision@1} where we compare the highest-ranked results from the 20 random samples 
against a user judgement of 1 or 2 (= strongly or mildly interested).
For completeness, we also consider {\bf NDCG@top10} 
for the top-10 results of the original,
potentially biased, rankings from a commercial search engine.

\vspace*{0.2cm}
\noindent{\bf Methods under Comparison:}
We 
cover
the following methods and configurations.
\squishlist
\item \textbf{LM} denotes the language model approach. To isolate the effect of the user model in the re-ranking, and as our initial pool of entities are to some extent relevant to the query, we either set the $\lambda$ to 0 or 1. When $\lambda=1$ the input to the re-ranker is the query model and when $\lambda=0$ only the user model is given as input.
\item\textbf{LM-embed}
is the language-model method with word embeddings using word2vec. The term-term similarity threshold is set to 0.5.
\item \textbf{BM25} is the BM25 method with 
parameters set to the following values
widely used in the literature:
$b=0.75, k_1=1.5$. 
\item \textbf{KNRM} 
is the neural ranker, with
the maximum query and document lengths set to 50 and 5000, respectively.
The terms for the query/user model are obtained by tf order, selecting the top 50 distinct terms.
Document terms are the top-5000 terms.
Models are trained on data per domain with 504, 772 and 806 assessments for book, food and travel, respectively.\\
As this training is fairly low-end,
we also study a variant {\bf KNRM-all}
where we combine all domains 
into a single training set with 2082 labeled samples.
We report on ten-fold cross-validation with 8, 1 and 1 folds for training, validation and test, respectively. %
\item \textbf{SE} is the initial ranking from a commercial search engine.
\squishend

\subsection{User Models: None vs. Chats vs. Questionnaires (RQ0 and RQ1)}
Table \ref{tab:results-rq1} shows the 
{NDCG@20} results for the influence of different user models.
The top part of Table \ref{tab:results-rq1}
gives the overall results across all domains
(averaged over the 113 user-query pairs).
The other parts show per-domain results.
The user models under comparison here are
query-only vs. questionnaires-based vs. chats-based. For the latter two, we varied
the specific setting by deriving models
from all available inputs regardless of the domains ($All$), using only general questionnaires or chats ($Gen$, see Section \ref{sec:datacollection}),
using only domain-specific inputs ($Dom$),
or using both general and per-domain inputs
($Dom+Gen$).
In this comparison, all methods were configured without entity expansion and without domain-specific vocabularies
(which will be discussed in the next subsections).

\input{chapters/tables/table-RQ1-significance}
\noindent{\bf Overall results (top part of Table \ref{tab:results-rq1}):}
The overriding observation is that almost all rankers with different degrees of personalization improve over the SE baseline %
and that both questionnaire-based and chat-based user models achieve notable gains over the query-only rankers: in the order of 2 to 4 percentage points in NDCG@20.
While the effect size of personalization is only moderate, the relative gains are
statistically significant and come at little cost for the ranker efficiency. 
For significance, two-tailed paired t-tests in
comparison to the Query-Only baselines mostly had p-values
$< 0.05$.
These results are marked with an asterisk in
Table \ref{tab:results-rq1}.

Interestingly, LM-embed did not improve over LM.
The term-term relatedness by word2vec seems to be too crude for our task and dilutes the query focus.
KNRM and KNRM-all were inferior to the Query-Only case.
The combination of small training data and limited input size is the likely cause for this disappointing result.

When comparing questionnaire-based vs.
chat-based personalization, the former
performs slightly better than the latter,
but the differences are minor.
For both, the best variants were the ones
with user models $Dom$ or $Dom+Gen$, indicating
awareness of the domain is beneficial.
$Dom$ is almost always preferable to $Dom+Gen$ in the case of chats, but there is no clear trend when using questionnaires.
This 
is likely due
to the fact that the 
general questionnaires
were 
designed
to reveal user personalities, whereas the general chats were mostly introductory and less informative.
These gains are not always statistically significant, but the best cases are: for example, the 
improvement for LM from 0.811 with chats-$All$ to 0.822 with chats-$Dom$ had a p-value of 0.0018.

\vspace*{0.1cm}
\noindent {\bf Per-domain results:}
The results vary  among the different domains in an interesting way.
We base the discussion on LM and BM25 as they achieved the best results.
For books and travel, the gains from 
user models are most pronounced.
For books, the chat-based models 
achieved a small but notable and significant
improvement over 
the questionnaire-based ones.
We observe that for questionnaire-based models $Dom+Gen$ outperforms $Dom$. This 
is due to the low coverage of the book domain with only two questions 
on the user's favorite books and genres, whereas the general questionnaire 
includes
demographics and personal traits.
On the other hand, for the travel domain with 5 specific questions, $Dom$ performs better than $Dom+Gen$ in both questionnaire-based and chat-based models, with the former giving the best results.

For food,
personalization
led to gains, but the absolute NDCG scores were substantially lower than for the 
other two domains.
Here, the SE performed better 
than the re-rankers with the query-only model. 
However, using $Dom+Gen$ questionnaire-based profiles,
we achieved up to 2\% improvement  over the SE results.
It seems that the food
domain is inherently difficult to
understand, as its vocabulary mixes
specific and very common words with
a strong influence of the latter on
tastes and sentiments (e.g., ``hot'',
``terrific'' etc.).

As for precision@1, the overall gains
by personalization were nearly 10 percent:
considering the best-performing rankers on overall results, the LM improved 
from 70\% with query-only models to
81\% with questionnaire-based models, %
and BM25 went up from 66\% to 83\%.
Again, the gains were most substantial
for books and travel, but here
food as well showed
notably improved precision@1. %
We further evaluated NDCG@top10:
not surprisingly, the SE baseline
was stronger for this metric, but was 
still outperformed by re-ranking with
personalization. The best values
for our method were comparable to
those for NDCG@20, around 83\%
across all domains and up to 87\%
for travel.

\subsection{Domain Vocabularies (RQ2)}

Recall from Section \ref{sec:domainspecificity}
that we optionally incorporate 
domain-specific term weighting to
reduce the influence of
irrelevant wording from the user chats.
Table \ref{tab:results-rq2} shows 
NDCG@20 results
with this awareness of domain vocabularies,
for the four chat-based 
configurations $All$, $Gen$, $Dom$ and
$Dom+Gen$. 
We show only overall results across all domains, but for each domain, all user-model terms were weighted by the respective
$spy(w)$ domain model.
For brevity, we restrict
ourselves to the LM-based ranker; the
findings were similar for the other two rankers.

\input{chapters/tables/table-RQ2}

Table \ref{tab:results-rq2} indicates
that there are small gains from this
domain-specific weighting, but the
effect size is marginal and not statistically significant (p-value $> 0.1$). 
It seems that chats are not sufficiently focused on domain-specific topics. Humans do jump between topics,
so chats naturally have a high level of thematic diversity.

\subsection{Entity Expansion (RQ3)}

To study the influence of entity expansion for the user models, we compared
different settings against the previously
reported configurations without entity
awareness:
{\em all} expands all entities including
concepts (in Wikipedia, such as
``history'' or ``Buddhist art'');
{\em domain} restricts the entities to
those that are related to the respective domain (see Section \ref{sec:entityexpansion});
{\em NE-all} uses only named entities
(i.e., discarding general concepts);
{\em NE-dom} uses only named entities
with domain relatedness above a threshold.

\input{chapters/tables/table-RQ3-significance}

Table \ref{tab:results-rq3} shows the
overall NDCG@20 for these settings
with the different configurations for
the user-model construction.
We observe that almost none of the expansion methods significantly improve the models
derived from questionnaires. %
The reason is
that these models are already very concise
given their high-quality inputs.
For chat-based user models, on the other hand,
entity expansion led to small, but notable and statistically significant 
(p-values 
$< 0.05$ ), 
improvements of ca. 1\%.

%% file: chapters/tables/table-RQ1-significance.tex
\begin{table}[b!]
\caption{NDCG@20 for different rankers and user models. 
Best results per row are in boldface.
Statistically significant improvements over the Query-Only baselines are marked with an asterisk.}
\centering
\small
\begin{tabular}{@{}l l  llll  llll@{}}
\toprule
Ranker$\phantom{12345}$ & \multicolumn{9}{c}{User Model} \\
\cmidrule(lr){3-10}
 & \multirow{2}{*}{\shortstack[l]{Query\\ Only}} &\multicolumn{4}{c}{Questionnaires\vspace{0pt}} &
\multicolumn{4}{c}{Chats\vspace{0pt}}\\
\cmidrule(lr){3-6} \cmidrule(lr){7-10}
 &&
{ All} & {Gen} & {Dom} & {Dom+Gen} &
{ All} & {Gen} & {Dom} & {Dom+Gen}
\\
\midrule\midrule
\Tstrut 
LM & 0.796 & 0.816 & 0.804 & 0.823$^{*}$ & \textbf{0.824}$^{*}$ & 0.811 & 0.806 & 0.822$^{*}$ & 0.817$^{*}$\\
LM-embed & 0.794 & 0.791 & 0.787 & \textbf{0.811} & 0.798 & 0.782 & 0.777 & 0.795 & 0.784\\
BM25 & 0.785 & 0.823$^{*}$ & 0.815$^{*}$ & 0.827$^{*}$ & \textbf{0.833}$^{*}$ & 0.819$^{*}$ & 0.816$^{*}$ & 0.827$^{*}$ & 0.821$^{*}$\\
KNRM & \textbf{0.807} & 0.791 & 0.805 & 0.798 & 0.794 & 0.780 & 0.786 & 0.784 & 0.785\\
KNRM-all & \textbf{0.810} & 0.807 & 0.796 & - & - & 0.788 & 0.791 & - & -\\
SE & 0.786  & - & - & - & - & - & - & -  \\
\bottomrule
\bottomrule
Books\\
\hline
LM & 0.825 & 0.829 & 0.823 & 0.822 & 0.834 & 0.846 & \textbf{0.854} & 0.844 & 0.847\\
LM-embed & \textbf{0.818} & 0.795 & 0.803 & 0.801 & 0.799 & 0.811 & 0.799 & 0.813 & 0.806\\
BM25 & 0.814 & 0.843 & 0.846 & 0.834 & 0.847 & 0.846 & 0.849 & \textbf{0.851} & 0.850\\
KNRM & 0.826 & 0.827 & \textbf{0.832} & 0.817 & 0.816 & 0.790 & 0.810 & 0.790 & 0.809\\
SE & 0.777 & - & - & - & - & - & - & - \\
\bottomrule
Travel\\
\hline
LM & 0.818 & 0.821 & 0.815 & \textbf{0.854}$^{*}$ & 0.838 & 0.826 & 0.813 & 0.841 & 0.835\\
LM-embed & 0.813 & 0.799 & 0.787 & \textbf{0.849}$^{*}$ & 0.814 & 0.785 & 0.782$^*$ & 0.803 & 0.796\\
BM25 & 0.794 & 0.837$^{*}$ & 0.833$^{*}$ & \textbf{0.857}$^{*}$ & 0.849$^{*}$ & 0.836$^{*}$ & 0.837$^{*}$ & 0.844$^{*}$ & 0.838$^{*}$\\
KNRM & \textbf{0.838} & 0.806 & 0.833 & 0.827 & 0.801 & 0.800 & 0.800 & 0.805 & 0.800\\
SE & 0.794  & - & - & - & - & - & - & - \\
\bottomrule
Food\\
\hline
LM & 0.753 & 0.802$^*$ & 0.779 & 0.790 & \textbf{0.803}$^*$ & 0.772 & 0.766 & 0.785 & 0.777\\
LM-embed & 0.757 & 0.778 & 0.775 & 0.777 & \textbf{0.780} & 0.758 & 0.755 & 0.773 & 0.757\\
BM25 & 0.756 & 0.793 & 0.775 & 0.791 & \textbf{0.806}$^*$ & 0.783 & 0.770 & 0.793$^*$ & 0.782\\
KNRM & 0.761 & 0.751 & 0.756 & 0.755 & \textbf{0.771} & 0.752 & 0.753 & 0.757 & 0.753\\
SE & 0.785 & - & - & - & - & - & - & - \\
\bottomrule
\end{tabular}
\label{tab:results-rq1}
\end{table}

%% file: chapters/tables/table-RQ2.tex
\begin{table}[!b]
\caption{NDCG@20 for LM-based ranker with domain-specific vocabularies}
\vspace*{-0.2cm}
\centering
\small
\begin{tabular}{@{}l cccc @{}}
\toprule
Domain Specificity $\phantom{12345}$ &  \multicolumn{4}{c}{Chats} \\
\cmidrule(lr){2-5} 
& { All} & {Gen} & {Dom} & {Dom+Gen} \\
\bottomrule
Disabled & 0.811 & 0.806 & 0.822 & 0.817\\
Enabled & 0.821 & 0.813 & 0.83 & 0.826\\
\bottomrule
\end{tabular}
\label{tab:results-rq2}
\end{table}

%% file: chapters/tables/table-RQ3-significance.tex
\begin{table}[!t]
\caption{NDCG@20 for LM-based ranker 
with entity expansion. Best results per column are in boldface.
Statistically significant improvements over 
{None} baselines
are marked with an asterisk.}
\vspace*{-0.2cm}
\centering
\small
\begin{tabular}{@{}l  llll  llll@{}}
\toprule
Entity Expansion & \multicolumn{8}{c}{User Models} \\
\cmidrule(lr){2-9}
 &\multicolumn{4}{c}{Questionnaires\vspace{0pt}} &
\multicolumn{4}{c}{Chats\vspace{0pt}}\\
\cmidrule(lr){2-5} \cmidrule(lr){6-9}
 &
{ All} & {Gen} & {Dom} & {Dom+Gen} &
{ All} & {Gen} & {Dom} & {Dom+Gen}
\\
\midrule\midrule
\Tstrut 
None & 0.816 & 0.804 & 0.823 & 0.824 & 
0.811 & 0.806 & 0.822 & 0.817\\

All & 0.817 & \textbf{0.816} & 0.826 & 0.822 & 
\textbf{0.821} & 0.814 & 0.828 & 0.824\\

Domain & 0.823 & 0.812 & \textbf{0.827} & 0.824 & 
\textbf{0.821}$^*$ & 0.814$^*$ & 0.829 & 0.824\\

NE-all & 0.823 & 0.81 & 0.826 & 0.829 & 
0.818$^*$ & 0.813 & 0.829 & \textbf{0.825}$^*$\\

NE-dom & \textbf{0.829}$^*$ & 0.809 & 0.825 & \textbf{0.833} & 
0.819$^*$ & \textbf{0.815}$^*$ & \textbf{0.83} & 0.824$^*$\\
\bottomrule
\end{tabular}
\label{tab:results-rq3}
\end{table}

%% file: chapters/relatedwork.tex
\section{Related Work}
{\bf Recommender systems} are ubiquitous
in search, e-commerce and social content sharing.
Most state-of-the-art systems learn from massive
amounts of user-behavior signals: queries, clicks,
likes, ratings, etc. 
(e.g., \cite{DBLP:conf/recsys/ZhaoHWCNAKSYC19,Lalmas:NetflixWorkshop,DBLP:conf/www/JiangWRCYCGJC020}).
To a lesser extent, product reviews are considered
as well (see, e.g., \cite{DBLP:journals/umuai/ChenCW15,DBLP:conf/kdd/Bauman0T17}), but recent studies
\cite{DBLP:conf/adbis/StratigiLSZ19,SIGIR2020:sachdeva:useful} 
indicate that there is considerable noise in user reviews and limited benefit from
such additional input.
In the opposite direction, \cite{DBLP:conf/sigir/BalogRA19} made the point
that user models for personalized recommendations
should be scrutable and, therefore, use as little information
as possible and make the derived models transparent and
user-interpretable. 
The work \cite{CHIIR20:PES} pursued this rationale
by building on explicit user profiles from short questionnaires.
The current paper's experiments include comparisons to
that approach. 
None of the prior works has considered user-user chats
as a source for capturing user interests and tastes.
Note that interactive and conversational recommenders \cite{DBLP:conf/sigir/SchnabelABBJ20,DBLP:conf/sigir/Lei0RC20} are a very
different approach, as they build on dialogs
between user and system, not among users.

{\bf Specialized recommenders} that tap
textual contents have been
investigated for domains like e-learning, literature
exploration or tourism 
(e.g., \cite{DBLP:series/isrl/2017-112,DBLP:journals/access/BaiWLYKX19,DBLP:journals/tois/LiCPR19,DBLP:journals/corr/abs-1903-12099}).
These are based on rich context models of user history
and interests. However, they are not query-based,
disregarding the additional component of search results
on behalf of the user.

{\bf Personalized ranking of search results}
has been addressed from two angles
(see \cite{Ghorab2013} for a survey):
i) building user models from user queries and browsing
histories (e.g., \cite{DBLP:conf/sigir/TeevanDH05,DBLP:conf/cikm/ShenTZ05,agichtein2006learning,DBLP:conf/sigir/ChiritaFN07,kuzi2017query}), and
ii) exploiting such models for ranking, query
expansion or auto-completion 
(e.g., \cite{DBLP:conf/wsdm/MatthijsR11,DBLP:conf/sigir/Shokouhi13,DBLP:conf/sigir/CaiR16}).
For the first task, the seminal work of 
\cite{DBLP:conf/sigir/TeevanDH05} analyzed user activities
reflected in queries, clicks and emails,
all the way to news and other contents read by a user.
For personalized ranking,
language models were enhanced with user-specific priors
\cite{DBLP:conf/wsdm/SontagCBWDB12}.
The interplay of long-term behavior and short-term sessions of a user
was studied by
\cite{DBLP:conf/sigir/BennettRWY11,DBLP:conf/sigir/BennettWCDBBC12}.
Other work \cite{DBLP:conf/sigir/TeevanDL08,DBLP:conf/ictir/BennettSC15} investigated
the issue of when to personalize and 
when to disregard user profiles.
None of these prior works is specifically geared
for entity search,
and none considers user models derived from chats.

{\bf Entity search} (e.g., \cite{DBLP:series/irs/Balog18,DBLP:conf/sigir/Dietz19})
has been studied for personalization only
in limited settings.
The CLEF competition on book recommendations
\cite{DBLP:conf/clef/KoolenBGHHHKSVW16}
relied on extensive data (posts, tags, reviews, ratings) by 
large user communities at
LibraryThing and Amazon.
Most related to our work is \cite{Ai:2017:LHE:3077136.3080813}
on personalized product search,
based on embeddings for users and products in
a joint latent space.
That method exploited user reviews on product pages. 
In contrast, our approach is based on user-user
chats, an unintrusively observable 
asset disregarded in prior works.

{\bf Query expansion} is a well established
methodology in IR (see, e.g., \cite{DBLP:journals/csur/CarpinetoR12} for a survey).
Personalization has been studied in this context
along various routes. Notable examples are
\cite{DBLP:journals/tist/BiancalanaGMS13a,zhou2017query} based on
user-provided tags,
and
\cite{kuzi2017query} based on email histories
and utilizing word embeddings learned from
email contents.
Recently, \cite{DBLP:conf/sigir/YaoDW20}
has pursued the theme of personalized
word embeddings further, based on
query histories.

%% file: chapters/conclusion.tex
\section{Conclusion}
To the best of our knowledge, this is the first work that 
explores leveraging 
{\em user-to-user} conversations as a source for personalization of search-based recommendations.
We compared chat-based user models
against models derived from concise questionnaires.
Both achieved substantial improvements over both the original search-engine ranking and non-personalized query-only re-rankings. 

Between chat-based and questionnaire-based re-rankings, there is no clear winner.
The two paradigms of user models each have specific benefits:
\squishlist
\item Questionnaries are 
transparent and scrutable for users. However, they require an explicit effort. Most users seem fine with a one-time questionnaire, but few seem ready for periodic updating as their interests and tastes evolve.
\item Chats, on the other hand, require no effort at all from the user side, and could be easily updated without user intervention.
However, the derived models are less
transparent to humans and not easily
adjustable by users themselves.
Also, chat data comes with higher privacy risks.
\squishend

\noindent 
The additional enhancements devised in this paper -- domain awareness and entity expansion -- further improved the NDCG scores, but only to a small extent. 
On the other hand, focusing on entities in conversations and casting them into
an explicit user model is a step towards making chat-based profiles more 
transparent and scrutable for users.